\documentclass
[twocolumn,showpacs,preprintnumbers,amsmath,amssymb]
{revtex4}
\usepackage{graphicx}
\usepackage{dcolumn}
\usepackage{bm}
\usepackage{here}
\usepackage{color}
\usepackage{braket}
\usepackage{amsmath}  
\usepackage{amssymb}  
\usepackage{fancybox}

\begin{document}

\title{Dynamical Localization and Delocalization in Polychromatically Perturbed 
Anderson Map}
\author{Hiroaki S. Yamada}
\affiliation{Yamada Physics Research Laboratory,
Aoyama 5-7-14-205, Niigata 950-2002, Japan}
\author{Kensuke S. Ikeda}
\affiliation{College of Science and Engineering, Ritsumeikan University
Noji-higashi 1-1-1, Kusatsu 525-8577, Japan}

\date{\today}
\begin{abstract}
In the previous paper [PRE 101,032210(2020)], localization and delocalization phenomena 
 in the polychromatically perturbed Anderson map (AM)
were elucidated mainly from the viewpoint of localization-delocalization transition
 (LDT) on the increase of the perturbation strength $\epsilon$. 
In this paper, we mainly investigate the dependency of the phenomena 
on the disorder strength $W$ ($W-$dependence)
 in the AM with a characteristic disorder strength $W^*$.
In the completely localized region the $W-$dependence and $\epsilon-$dependence 
of the localization length show characteristic behavior similar to 
those reported in monochromatically perturbed cases [PRE 97,012210(2018)].
Furthermore,  the obtained results show that even for the increase of the $W$, 
the critical phenomenon and  
critical exponent are found to be  similar to those in the LDT caused by the increase of  $\epsilon$.
We also investigate the diffusive properties of the delocalized states 
induced by the parameters.
\end{abstract}

\pacs{05.45.Mt,71.23.An,72.20.Ee}


\maketitle


\newcommand{\vc}[1]{\mbox{\boldmath $#1$}}
\newcommand{\fracd}[2]{\frac{\displaystyle #1}{\displaystyle #2}}
\newcommand{\red}[1]{\textcolor{red}{#1}}
\newcommand{\blue}[1]{\textcolor{blue}{#1}}
\newcommand{\del}{\partial}

\def\ni{\noindent}
\def\nn{\nonumber}
\def\bH{\begin{Huge}}
\def\eH{\end{Huge}}
\def\bL{\begin{Large}}
\def\eL{\end{Large}}
\def\bl{\begin{large}}
\def\el{\end{large}}
\def\beq{\begin{eqnarray}}
\def\eeq{\end{eqnarray}}

\def\eps{\epsilon}
\def\th{\theta}
\def\del{\delta}
\def\omg{\omega}

\def\e{{\rm e}}
\def\exp{{\rm exp}}
\def\arg{{\rm arg}}
\def\Im{{\rm Im}}
\def\Re{{\rm Re}}

\def\sup{\supset}
\def\sub{\subset}
\def\a{\cap}
\def\u{\cup}
\def\bks{\backslash}

\def\ovl{\overline}
\def\unl{\underline}

\def\rar{\rightarrow}
\def\Rar{\Rightarrow}
\def\lar{\leftarrow}
\def\Lar{\Leftarrow}
\def\bar{\leftrightarrow}
\def\Bar{\Leftrightarrow}

\def\pr{\partial}

\def\>{\rangle} 
\def\<{\langle} 
\def\RR {\rangle\!\rangle} 
\def\LL {\langle\!\langle} 
\def\const{{\rm const.}}

\def\e{{\rm e}}

\def\Bstar{\bL $\star$ \eL}

\def\etath{\eta_{th}}
\def\irrev{{\mathcal R}}
\def\e{{\rm e}}
\def\noise{n}
\def\hatp{\hat{p}}
\def\hatq{\hat{q}}
\def\hatU{\hat{U}}

\def\hatA{\hat{A}}
\def\hatB{\hat{B}}
\def\hatC{\hat{C}}
\def\hatJ{\hat{J}}
\def\hatI{\hat{I}}
\def\hatP{\hat{P}}
\def\hatQ{\hat{Q}}
\def\hatU{\hat{U}}
\def\hatW{\hat{W}}
\def\hatX{\hat{X}}
\def\hatY{\hat{Y}}
\def\hatV{\hat{V}}
\def\hatt{\hat{t}}
\def\hatw{\hat{w}}

\def\hatp{\hat{p}}
\def\hatq{\hat{q}}
\def\hatU{\hat{U}}
\def\hatn{\hat{n}}

\def\hatphi{\hat{\phi}}
\def\hattheta{\hat{\theta}}

\def\iset{\mathcal{I}}
\def\fset{\mathcal{F}}
\def\pr{\partial}
\def\traj{\ell}
\def\eps{\epsilon}
\def\U{\hat{U}}

\def\U{U_{\rm cls}}
\def\P{P_{{\rm cls},\eta}}
\def\traj{\ell}
\def\cc{\cdot}

\def\DZ{D^{(0)}}
\def\Dcls{D_{\rm cls}}

\newcommand{\relmiddle}[1]{\mathrel{}\middle#1\mathrel{}}

\section{Introduction} 
In recent years, the localization of wave packet in quantum map systems has been 
extensively studied experimentally \cite{chabe08,wang09} and 
 theoretically \cite{lemarie10,tian11,lopez12,lopez13}.
Experimentally, Chabe {\it et al.} observed the critical phenomena of the 
localization-delocalization transition (LDT) for cold atoms in an optical lattice, 
 which corresponds to perturbed standard map (SM). 
The results are interpreted based on the equivalence to Anderson transition 
in the three-dimensional disordered tight-binding system \cite{ishii73,lifshiz88}.

\begin{figure}[htbp]
\begin{center}
\includegraphics[width=8.0cm]{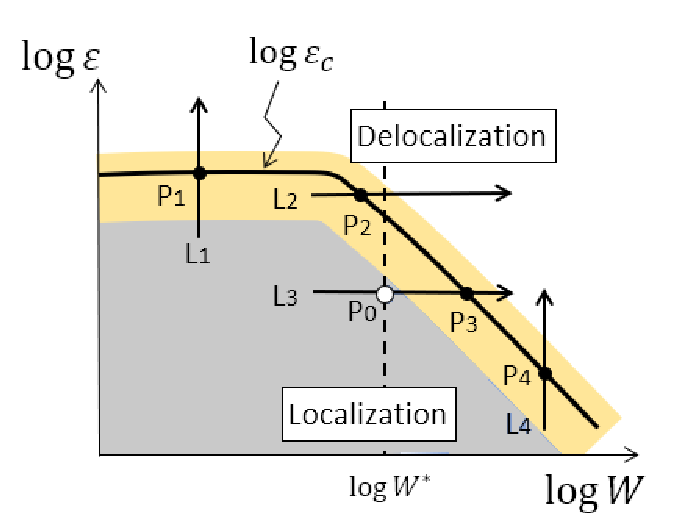}
\caption{(Color online)
The illustrating of the critical curve $\eps_c(W)$ in the 
phase plane $(\log \eps, \log W)$ 
for the dichromatically perturbed AM ($M=2$), 
where $W=W^*$ is shown by the dotted black line. 
Yellow region represents the critical region of LDT, 
gray represents the region where the localization length obey $W^{-2}$ law.
Some typical paths from the localized state to the delocalized state 
via the LDT due to parameter changes are represented by $L_n(n = 1, 2, 3, 4)$. 
$P_n(n = 1,2,3,4)$ represents the transition points in each case, and  
$ P_0 $ indicates the intersection with $W^*$ when $W$ is increased 
along $L_3$.
If the vertical axis is replaced to $\eps(M-1)$, the similar phase diagram
 holds for the polychromatically perturbed cases ($M >2$) \cite{yamada19a}.
}
\label{fig:phase-1}
\end{center}
\end{figure}

We have proposed Anderson map (AM) that becomes the one-dimensional
 Anderson model 
in a certain limit and investigated the parameter dependence of 
the LDT  in the AM with the time-quasiperiodic perturbation
composed of $M-$color modes \cite{yamada15,yamada19a}. 
Let us define $W$ as the disorder strength 
and $\eps$ as the perturbation strength, respectively.
The characteristic $\eps-$dependence and the $W-$dependence for the LDT 
in the perturbed AM
was clarified in comparison with the SM under the same perturbation.
A schematic representation of the resulting phase diagram for the LDT of 
the dichromatically perturbed AM ($M=2$) is shown in Fig.\ref{fig:phase-1}.
The critical curve $\eps_c(W)$ is shown as the $\log \eps$ on the vertical axis 
and the $\log W$ on the horizontal axis.

In the previous paper \cite{yamada15,yamada19a}, 
we investigated the LDT with  increasing 
the perturbation strength $\eps$
along the line $L_1$ or line $L_4$ in Fig.\ref{fig:phase-1}, and 
obtained the parameter dependence of the critical exponent of the localization length 
and critical strength $\eps_c$
by using finite-time scaling analysis of the LDT. 
Roughly speaking, the $W-$dependence of the critical curve $\eps_c(W)$
greatly changes  around $W \simeq W^*$.
Table \ref{table:transition} summarizes the difference
between  the time-evolution of the initially localized wave packet
in $W<W^*$ and $W>W^*$ when $\eps$ changes along $L_1$ and $L_4$ 
passing the LDT points $\eps=\eps_c$.

\begin{table*}[htbp]
\begin{tabular}{lll}
\hline \hline
             & $\eps <\eps_c$ & Ballistic $\to$ (Diffusive ) $\to$ Localization \\ \cline{2-3}
$W<W^*$ & $\eps =\eps_c$ & Ballistic $\to$ (Diffusive ) $\to$Subdiffusion \\ \cline{2-3}
  ($L_1$)  & $\eps >\eps_c$ & Ballistic $\to$ (Diffusive ) $\to$ Diffusion \\ \hline \hline
             & $\eps <\eps_c$ & (Ballistic) $\to$ Diffusive $\to$ Localization \\ \cline{2-3}
$W>W^*$ & $\eps =\eps_c$ & (Ballistic) $\to$ Diffusive $\to$ Subdiffusion \\ \cline{2-3}
  ($L_4$)   & $\eps >\eps_c$ & (Ballistic) $\to$ Diffusive $\to$ Diffusion \\ \hline \hline
\end{tabular} 
\caption{
The image of time-evolution of the spread for the initially localized wave packet
in the order of the arrows.
The transition from localization to diffusion is shown 
divided into two regions, $W<W^*$ and $W>W^*$,  corresponding 
the Fig.\ref{fig:c2c3-e005-start} and Fig.\ref{fig:W2-c4c6-long}, respectively.
$(...)$ represents the behavior observed in the very short time domain.
}
\label{table:transition}
 \end{table*} 

The purpose of this paper is to give some new results and 
data complementary to the previous paper
that have not been clearly shown yet for the localization and/or delocalization 
phenomena in the polychromatically perturbed AM ($M\geq2$).

The first is the localization characteristics in the region 
where $\eps$ is small ($\eps<\eps_c$) and $W$ is also  small.
In the paper \cite{yamada18}, we have already investigated the $W-$dependence of 
the localization length (LL) of a monochromatically perturbed AM.
As a result, the characteristics changes around $W\simeq W^*$.
In this paper, we investigated changes in the LL depending on $W$ and $\eps$ 
in the fully localized region
(corresponding to the gray region in Fig.1). 
For both regions, $W <W^*$ and $W> W^*$, 
the LL shows an exponential increase with concerning $\eps$, 
similar  to the result already reported in Ref.\cite{yamada18}.
When $W <W^*$, the LL shows the $W^{-2}-$decay, 
but for the region of $W>W^*$, it can be seen that 
the LL increases as the disorder strength $W$ increases, regardless of 
the number of the modes $M(\geq2)$.

Second, in this paper, we show  characteristic behavior of this system regarding LDT
when the perturbation strength $\eps$ is fixed, 
and the disorder strength $W$ is varied along the line $L_2$ or line $L_3$.
Let the critical disorder strength $W_c$. 
In particular, 
in the case of $W^* <W_c$, there is a change in the localization process 
(transient region of time toward the  localization) at $W=W^*$ before the 
LDT occurs at $W=W_c$, even if LDT finally occurs due to the increase of $W$.

Finally, we show the diffusive property of the delocalized states
for $\eps>\eps_c$ and/or $W>W_c$ 
by using the diffusion coefficient $D$.
The diffusion coefficient $D$ behaves as $D \propto W^{-2}$ for $W<W^*$, 
and it makes minimum diffusion coefficient around $W\simeq W^*$
and it gradually increases towards a constant value for $W>>W^*$ regardless of $M(\geq2)$.
The $W-$dependency for the $W >>W^*$ region can be explained 
by the ballistic spreading without localization even for the unperturbed case.

The organization of this paper is as follows.
In the next section, we introduce the perturbed Anderson map 
with quasiperiodic modulation and the Maryland transform \cite{grempel84}.
We show the localization property of the system
in the Sect.\ref{sec:exp-scaling}.
The $W-$dependence of  LDT is shown in the Sect.\ref{sec:AM-critical-MSD}.  
In the Sect.\ref{sec:delocalized-state}, we show the diffusive property of
the delocalized states.

Also, considering the LDT along the line L1 in Fig.\ref{fig:phase-1}, 
we consider the relationship between the LL in the yellow critical region 
and the LL in the grey region (strong localization) in  appendix A.
In addition, we investigated how polychromatic perturbation affects 
quantum states in ballistic spreading without localization in appendix B.

\section{Models} 
\label{sec:model}
The time evolution from $m$th step to $(m+1)$th step for the wave packet
$|\Psi>$ is described by 
\beq
|\Psi(m+1)>=\hat{U}_m|\Psi(m)>.
\eeq
The one-step time-evolution operator of  the following Anderson map is
\begin{eqnarray}
\label{eq:tight-binding}
 \hat{U}_m= 
\e^{-if(m)Wv(q)/\hbar} 
\e^{-iT(p)/\hbar},
\end{eqnarray}
where $T(p)=2\cos(p/\hbar)=\e^{-d/dq}+\e^{+d/dq}$ 
 is the kinetic energy term
and 
$v(q)=\sum_n\delta(q-n) v_q|q><q|$
is random on-site potential.
Here $\hatp$ and $\hatq$ are momentum and position
operators, respectively. 
The $v_n$ is uniformly distributed 
over the range $[-1,1]$, and  $W$ denotes the disorder strength. 
It is a quantum map version
of the Anderson model defined on the discretized
lattice $q \in {\Bbb Z}$ \cite{yamada99}. 
The quasiperiodic modulation $f(t)$ is given as,
\beq
f(t) =\left[1+ \frac{\eps}{\sqrt{M}} \sum_j^M \cos(\omega_j t)\right] 
\label{eq:perterbation}
\eeq
where $M$ and $\eps$ are number of the frequency component and 
the strength of the perturbation, respectively.
Note that the strength of the perturbation is divided by $\sqrt{M}$
to make the total power of the long-time average 
independent of $M$, i.e. $\overline{f(t)^2}=1+\eps^2/2$, and 
the frequencies $\{ \omega_i\}(j=1,...,M)$ are taken 
as mutually  incommensurate number of $O(1)$.

We can regard the harmonic perturbation as the dynamical degrees of freedom. To show this
we introduce the classically canonical action-angle operators $(\hatJ_j=-i\hbar \frac{\pr_j}{\pr_j\phi_j}, \phi_j)$
representing the harmonic perturbation as the linear modes and we 
 call them the ``color modes'' hereafter.
We consider the Hamiltonian $H_{aut}$ so as to include the color modes,
\beq
&& H_{aut}(\hatp,\hatq,\{\hatJ_j\},\{\hatphi_j\}) =T(\hatp)+ \nn \\
&& 
Wv(\hatq) \left[1+ \frac{\eps}{\sqrt{M}} \sum_j^M \cos \phi_j \right]\delta_t
+\sum_{j=1}^M \omega_j\hatJ_j, 
\eeq
where  $\delta_t=\sum_{m=-\infty}^{\infty}\delta(t-m)$.
One can easily check that by Maryland transform
the eigenvalue problem of the quantum map system interacting with 
$M$-color modes can be transformed into $d(=M+1)$-dimensional
lattice problem with the disorder. 
Let us consider an eigenvalue equation 
\beq
  \e^{-i\hatA}\e^{-i\hatB}\e^{-i\hatC}|u\> =\e^{-i\gamma}|u\>, 
\label{eq:eigen-value-problem}
\eeq
where 
\beq
\begin{cases}
\hatA =(Wv(\hatq)+\sum_j^M\omega_i \hatJ_j)/\hbar , \\
\hatB =v(\hatq)\frac{\eps W}{\sqrt{M}}\sum_j^M\cos\phi_j/\hbar , \\
\hatC =2\cos (\hatp/\hbar)/\hbar 
\end{cases}
\eeq
for the time-evolution operator.
$\gamma$ and $|u\>$ are the quasi-eigenvalue and quasi-eigenstate.
Here, if the eigenstate representation of $\hatJ_j$ is used,
$\hatJ_j|m_j\>=m_j\hbar|m_j\>$($m_j\in {\Bbb Z}$), 
we can obtain the following 
$(M+1)-$dimensional tight-binding expression 
by the Maryland transform \cite{yamada19a}:
\beq
\label{eq:Maryland_AM}
& & D(n,\{m_j\})u(n,\{m_j\}) +  \nn \\
& & \sum_{n',\{m_j^{'}\}}\<n,\{m_j\}|\hat{t}_{AM}|n',\{m_j^{'}\}\>  
u(n',\{m_j^{'}\}) =0, 
\eeq
where $\{m_j\}=(m_1,....,m_M)$.
Here the diagonal term is 
\beq
D(n,\{m_j\})=
\tan \left[ \frac{Wv_n+\hbar\sum_j^Mm_j\omega_j}{2\hbar}-\frac{\gamma}{2}
 \right], 
\eeq
and the $\hat{t}_{AM}$ of the off-diagonal term is 
\beq
\hat{t}_{AM}=i\frac{
e^{-i  \frac{\eps W}{\sqrt{M}}v(\hatq) (\sum_{j}^{M}\cos\phi_j)/\hbar}-
e^{i2\cos(\hatp/\hbar)/\hbar}
}
{e^{-i\frac{\eps W}{\sqrt{M}}v(\hatq) (\sum_{j}^{M}\cos\phi_j) /\hbar}+
e^{i2\cos(\hatp/\hbar)/\hbar}
}.
\eeq
It follows that the $(M+1)-$dimensional tight-binding models
of the AM have singularity of the on-site energy caused by 
tangent function and long-range hopping caused by the kick $\delta_t$.

If the off-diagonal term does not change qualitatively, there exists 
$W^*=2\pi\hbar \simeq 0.78$ where the effect of the
the fluctuation width of the diagonal term is saturated
for the change of  the disorder strength $W$.
It can be seen that for $W> W^*$ the diagonal fluctuation width is saturated,
and the effect of the $W$ is effective only for hopping term in the form of $\eps W$.
Therefore, it is suggested that for $W> W^*$ the phenomenon related to the transition 
phenomenon can be scaled in the form of $\eps W$.

\section{Localization properties in the polychromatically perturbed Anderson map}
\label{sec:exp-scaling}
We use an initial quantum state $<n|\Psi(t=0)>=\delta_{n,N/2}$ 
and monitor the spread of the wave packet by the 
mean square displacement (MSD),
\beq
  m_2(t)=<\Psi(t)|(\hat{n}-N/2)^2|\Psi(t)>.
\eeq
In the unperturbed cases  ($\eps=0$), it is known that 
the AM shows the localization in the real lattice space and 
the localization can be retained 
even when the monochromatic perturbation mode is added 
($\eps \neq 0 $, $ M = 1 $).
We have reported that in the monochromatically perturbed case ($M=1$)
the $W^{-2}$-dependence of the  localization length (LL)
 is stably maintained even for $\eps \neq 0$
and $W<W^{*}$, but the LL increases with increase of $W$ for $W>W^*$
at least in the weak perturbation limit $\eps<<1$.
In the case with $M \geq 2$, the LDT occurs by increasing $\eps$ and $W$.
However, the localization characteristics in the polychromatically perturbed AM
with small $\eps(<\eps_c)$  have not been investigated yet.
The purpose of this section is to show this.

\subsection{Localization length}
\begin{figure}[htbp]
\begin{center}
\includegraphics[width=8.5cm]{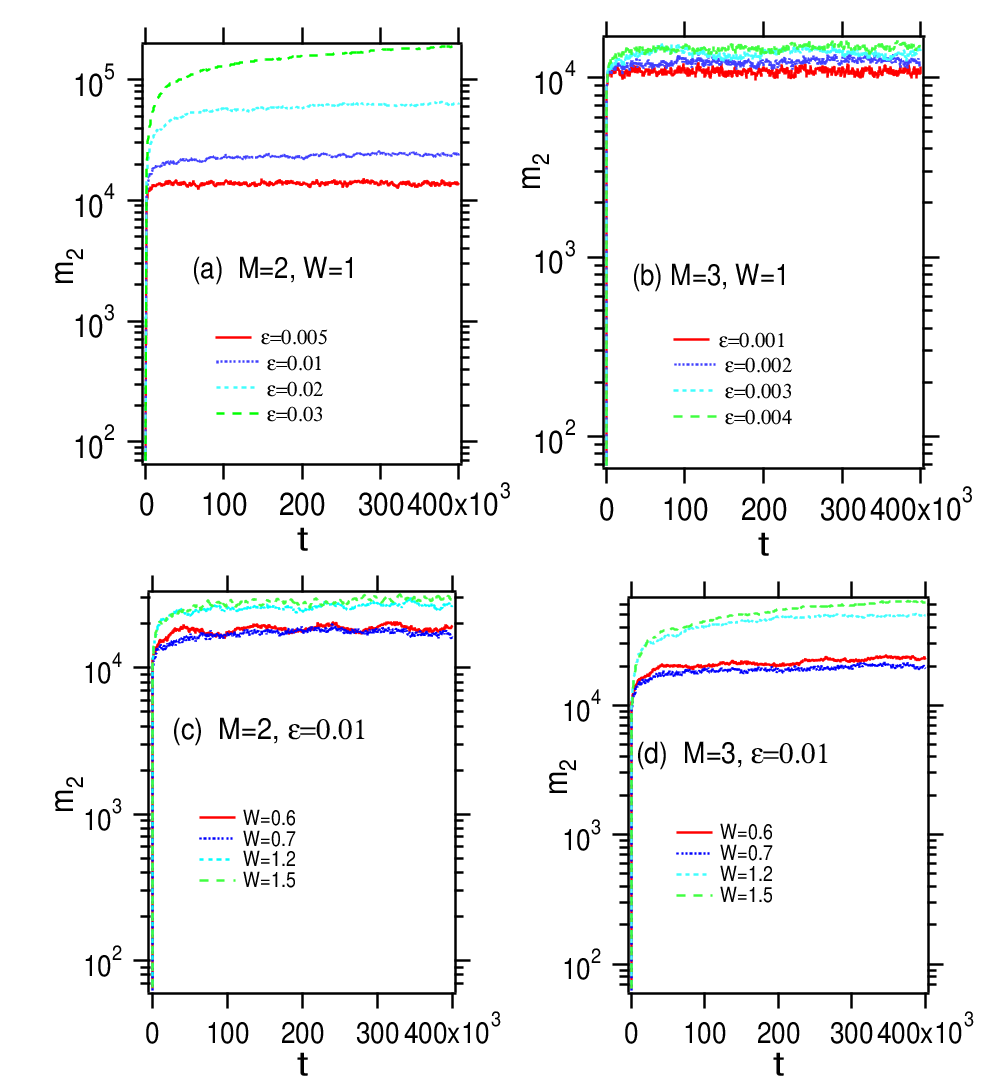}
\caption{(Color online)
The plots of $m_2(t)$ 
as a function of time
for different values of the $\eps$ and $W$
in the polychromatically perturbed AM.
 (a)$M=2$, $W=1.0$, (b)$M=3$, $W=1.0$, 
 (c)$M=2$, $\eps=0.01$ and  (d)$M=3$, $\eps=0.01$.
Note that the horizontal axes are in the logarithmic scale.
}
\label{fig:msd-1}
\end{center}
\end{figure}

We compute the LL of the dynamical localization, 
$\xi=\sqrt{m_{2}(\infty)}$ for the polychromatically perturbed cases,  
after numerically calculating the MSD for long-time,  
where $m_{2}(\infty)$ is numerically saturated MSD.
Figure \ref{fig:msd-1} shows time-dependence of the MSD 
for different values of the perturbation strength $\eps$ and the disorder strength $W$
when $M=2,3$.
It can be seen that the LL increases as $\eps$ increases when the $W$ is fixed,
and the  LL for the cases when the $\eps$ is fixed 
behaves somewhat complicated because of the existence of the $W^*$.

\begin{figure}[htbp]
\begin{center}
\includegraphics[width=7.0cm]{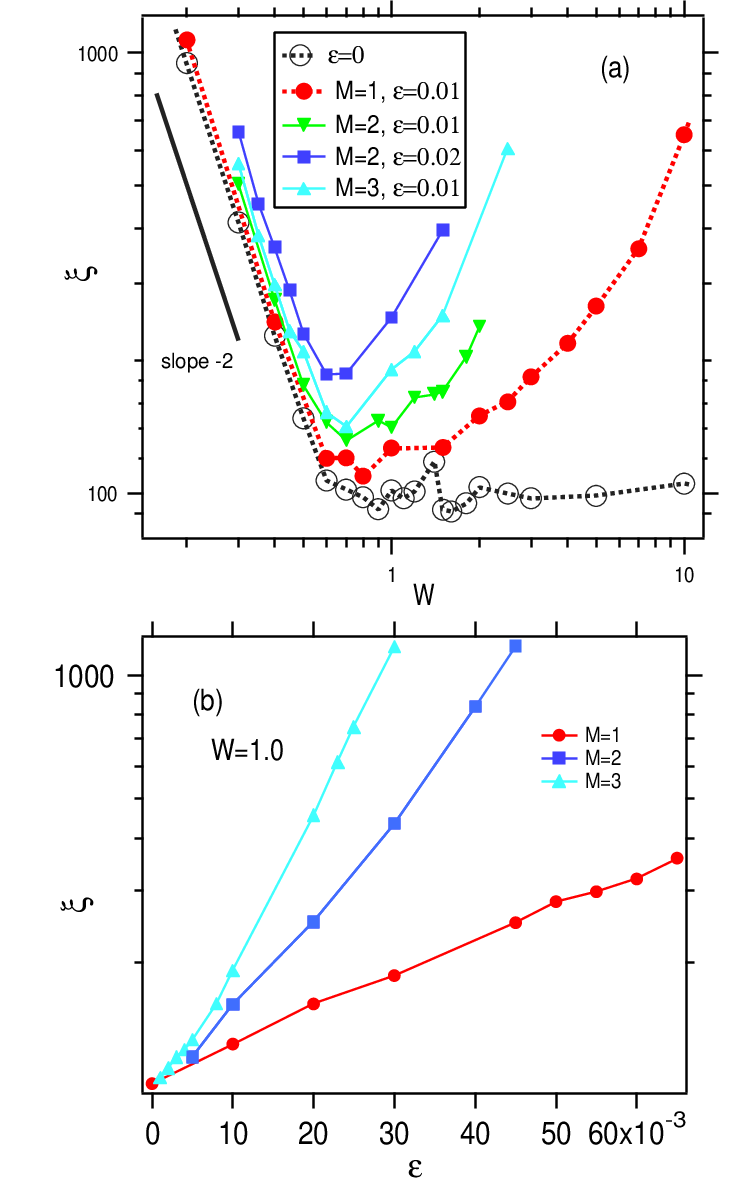}
\caption{(Color online)
(a)Localization length of the polychromatically perturbed AM 
as a function of disorder strength $W$ for various values of $M$ and $\eps$.
The unperturbed case ($\eps=0$) and monochromatically perturbed case  ($\eps=0.01$) 
 are also plotted by the dotted line with circles and filled circles, respectively.
Note that the axes are in the logarithmic scale.  
(b)Localization length of some perturbed AM ($M=1,2,3$) as a 
function of $\eps$ with $W=1.0$.
}
\label{fig:AM-LL-W-1}
\end{center}
\end{figure}

The $W-$dependence of the LL is over-plotted in Fig.\ref{fig:AM-LL-W-1}(a)
for $M=2$ and $M=3$.
We divid the $W$ region  into 
two regions, i.e., $W<W^{*}$ and $W>W^{*}$, to clarify their characteristics.
It follows that for $W \leq W^*$, the LL shows $W^{-2}-$decays like the case
of $M=1$, and it increases with respect to $W$ in the region $W>W^*$.

\begin{figure}[htbp]
\begin{center}
\includegraphics[width=8.5cm]{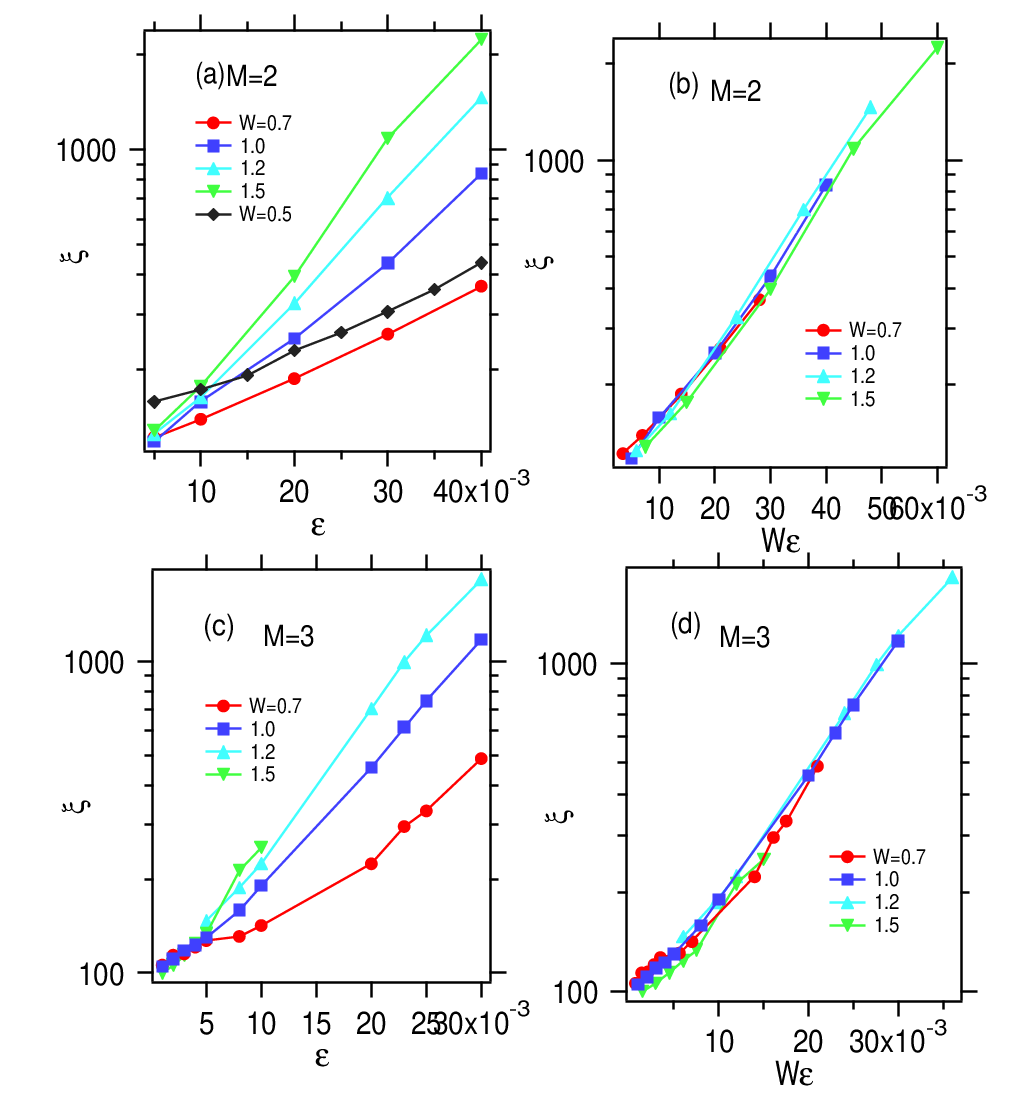}
\caption{(Color online)
(a)Localization length  $\xi$ of the dichromatically perturbed AM ($M=2$)
as a function of perturbation strength $\eps$ for 
the relatively large $W$.
(b)The plot of   $\xi$ as a function of $\eps W$ for  the panel (a).
(c)Localization length  $\xi$ of the dichromatically perturbed AM ($M=3$)
as a function of perturbation strength $\eps$.
(d)The plot of   $\xi$ as a function of $\eps W$ for  the panel (c).
Note that all the vertical axes are in the logarithmic scale. 
}
\label{fig:AM-LL-scale-2}
\end{center}
\end{figure}

Figure \ref{fig:AM-LL-W-1}(b) shows the result of 
the $\eps-$dependence in the the perturbed AM ($M=1,2,3$) for $W=1.0(>W^*)$.
It is obvious that the LL grows exponentially as 
the perturbation strength $\eps$ increases in all cases, i.e., 
\beq
 \xi \sim e^{c\eps},
\eeq
where $c$ is a growth rate.
Furthermore,  as shown in Fig.\ref{fig:AM-LL-scale-2}(a),(c), 
the exponential growth of the LL $\xi(\eps)$ is also confirmed by changing 
the disorder strength $W$.
Figures \ref{fig:AM-LL-scale-2}(b) and (d) show the plots of the 
(a) and (c) on the horizontal axis scaled as $\eps \to \eps W$, respectively.
We can see that it stands very well in one straight curve.
As a result, regardless of the number of color modes $M$ 
the parameter dependence of the localization length for $W>W^*$
is represented as
\beq
 \xi \sim e^{c_1\eps W}.
\eeq
This form is also the same as in the monochromatically perturbed AM.
It follows that the LL for $W<W^*$ behaves $\xi \sim e^{c_0\eps}$ with 
$W-$independent coefficient $c_0$ as shown in Fig.\ref{fig:AM-LL-scale-2}(c).

Therefore, if $\eps <<1$,
the $\eps-$dependence of the LL $\xi(\eps)$ of the polychromatically, perturbed cases
 also increase exponentially as:
\beq
\xi(\eps,W) \simeq \left\{
\begin{array}{ll}
c W^{-2} exp \{ c_0\eps \} & (W<W^*) \\
\xi_0^{*} exp \{ c_1 W \eps \} & (W>W^*),
\end{array}
\right.
\label{eq:AM-last}
\eeq
where the $c_0$ and $c_1$ are coefficients that increase with $M$.

Note that for the larger $\eps$, 
the LL $\xi(\eps)$ seems to increase more strongly than the exponential function 
in the cases of $M \geq 2$ in Fig.\ref{fig:AM-LL-W-1}(b).
This is a sign that leads to divergence at the critical point $\eps_c$
of the LDT.
See appendix \ref{app:LL-connect} for the more details.

\subsection{Scaling of the dynamical localization}
In this section, we recheck one-parameter scaling for the transient 
process to the localization.
For the $W<W^*$ region and  $W>W^*$ region,  
the different scaling property is shown. 

\begin{figure}[htbp]
\begin{center}
\includegraphics[width=4cm]{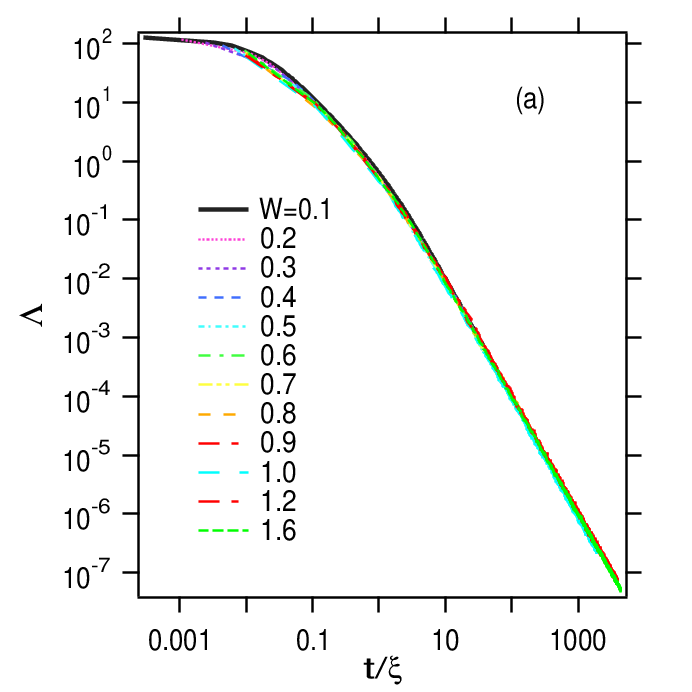}
\hspace{2mm}
\includegraphics[width=4cm]{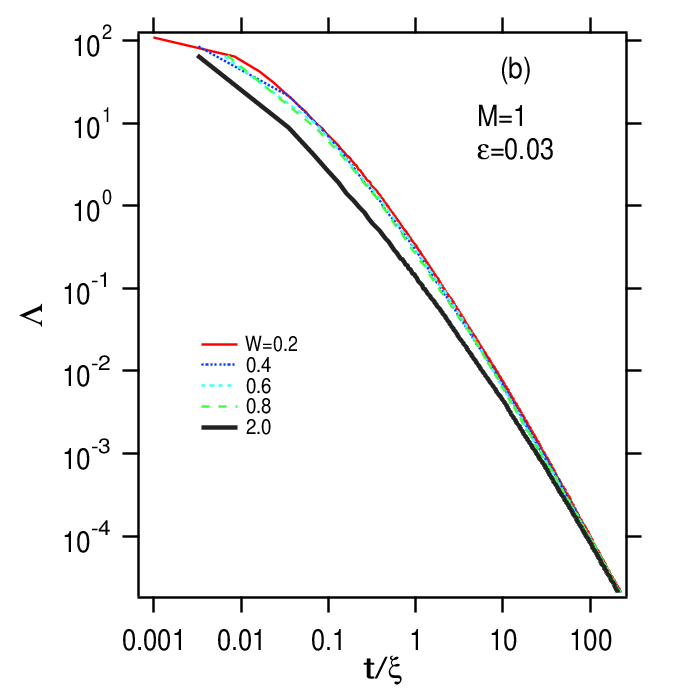}
\hspace{5mm}
\includegraphics[width=4cm]{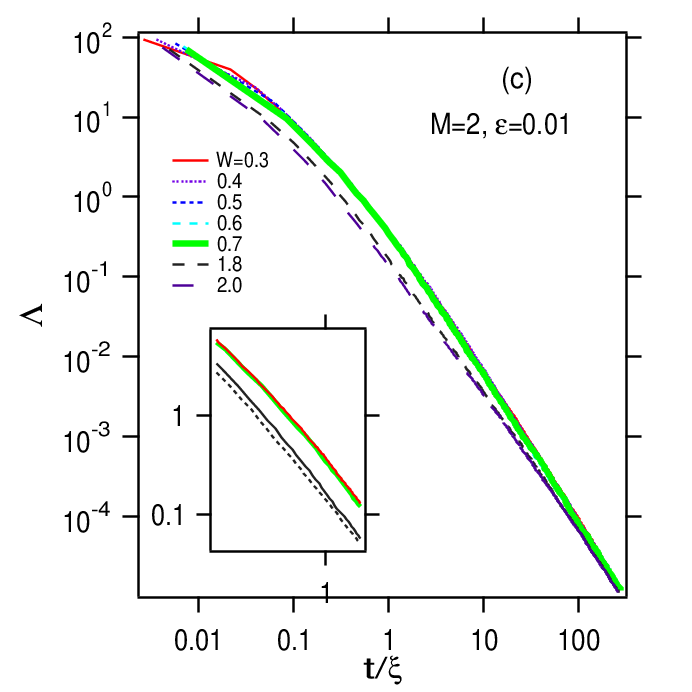}
\hspace{2mm}
\includegraphics[width=4cm]{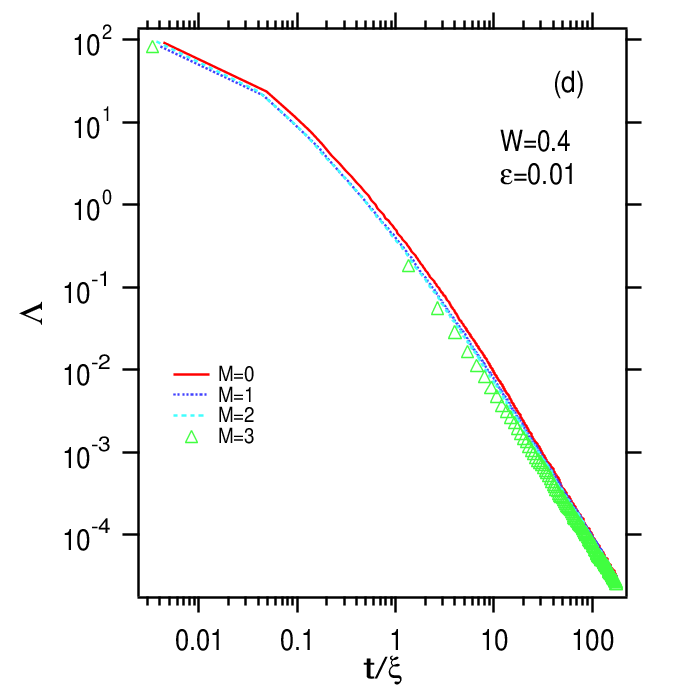}
\caption{
(Color online)
Scaled MSD $\Lambda(t)$ as a function of the scaled time
$t/\xi(W,\eps)$, where $\xi(W,\eps)$ are determined by MSD
for several parameter sets $(M, W, \eps)$.
(a)Unperturbed time-continuous Anderson model ($\eps=0$) with several values of the $W$.
(b)Monochromatically perturbed cases  ($M=1$, $\eps=0.03$) with several values of the $W$.
(c)Dichromatically perturbed cases  ($M=2$, $\eps=0.03$) with several values of the $W$.
(d)Perturbed cases  ($M=0,1,2,3$) of  the case with $W=0.4$ and $\eps=0.01$.
Note that all axes are in logarithmic scale.
}
\label{fig:scale-1}
\end{center}
\end{figure}

Time-dependence of the MSD shows ballistic growth $m_2 \sim t^2$ at the initial stage 
of the time-evolution for $W<W^*$, and it localizes $m_2 \sim t^0$ as $t \to \infty$.
Indeed, the ballistic spread of the wave packet 
has been observed for the short-time behavior of $m_2(t)$, as shown in
Fig.\ref{fig:c2c3-e005-start}(a).
We use the following scaled MSD as the scaling function:
\begin{eqnarray}
\Lambda(t)\equiv\frac{m_{2}(t)}{t^2} =F\left( \frac{t}{\xi}\right).
\label{eq:scale-1}
\end{eqnarray}
In this case, the fact that the localization process is scaled by the one-parameter 
$\xi$ in the entire time region means the following asymptotic form of 
the scaling function $F(x)$:
\beq
F(x) \sim       
  \begin{cases}
    const     & x \to 0\\
    \frac{1}{x^2}  & x \to \infty.
  \end{cases}
\label{eq:scaling-2}
\eeq
This type of scaling analysis has been performed 
to investigate LDT phenomena at the critical point
for polychromatically perturbed disordered systems \cite{chabe08}.
The establishment of one-parameter scaling means that the asymptotic shape is 
smoothly connected by a single curve if the localization length is used 
even in the case of various parameters.
Figure \ref{fig:scale-1}(a) shows the typical result of the localization phenomenon 
in (time-continuous) Anderson model 
by scaling the time with the localization length for 
various disorder strength $W$ determined by the numerical data of MSD.
It follows that the $\Lambda(t)$ roughly overlaps from the ballistic to the localized region.
The horizontal axis $x=t/\xi \to \infty$ shows a decrease of the slope $-2$.
These results suggest that the localization process 
from $m_2 \sim t^2$ to $m_2 \sim t^0$ is scaled by one-parameter 
when the localization occurs in the system.

Let examine the scaling for the perturbed AM.
Figure \ref{fig:scale-1}(b) and (c) show the result of applying the same scaling 
to the cases with  $M=1$ and $M=2$ with small $\eps$.
The localized states are obtained by changing various $W$s, 
and $\Lambda$ as a function of $t/\xi$ are displayed.
Since the $x \to \infty$ side is localized, it is obvious that it is asymptotic to 
$\Lambda_s(x) \sim x^{-2}$.
Actually, the data  with varying $W$ are also plotted, but the scaling is not bad.
However, if you look closely, it follows that 
the scaling works well for the cases for $W <W^*$,  
but we can see  the shift to the different curves  for $W>W^*$
in the transient region to the localization. 
(The inset in the Fig.\ref{fig:scale-1}(c) is an enlarged view.)
In other words, this is an existence effect of $W^*$, and when $W$
 is increased along the line $L_2$ in Fig.1, these features occur at the point $P_0$ 
denoted by the white circles.
In addition, as shown in the Fig.\ref{fig:scale-1}(d), 
it can be seen that  even for different $M$  
the similar scaling curves are obtained for  the region $W <W^*$.

\section{Localization-Delocalization transition in the polychromatically 
perturbed AM}
\label{sec:AM-critical-MSD}
In the previous paper \cite{yamada19a}, we analyzed in detail LDT caused with 
increasing $\eps$ for a fixed $W$ along Line $L_1$ in Fig.1.
In this section, we confirm the LDT by changing the disorder strength $W$
for the fixed $\eps$
along with the $L_2$ or $L_3$ in Fig.1.

\subsection{subdiffusion of LDT}
We can numerically determine the critical value $\eps_c$ and/or 
$W_c$ of the LDT so that the MSD becomes subdiffusion, 
\beq
m_2 \sim t^\alpha (0<\alpha<1).
\eeq
It is known from the numerical calculation that the diffusion index $\alpha$ is determined 
by the number of colors $M$ as, 
\beq
 \alpha \simeq \frac{2}{M+1},
\eeq
regardless of the LDT produced by changing $\eps$ 
or by changing $ W $.
This is  consistent with the prediction due to the one-parameter scaling (OPS).

\begin{figure}[htbp]
\begin{center}
\includegraphics[width=7.0cm]{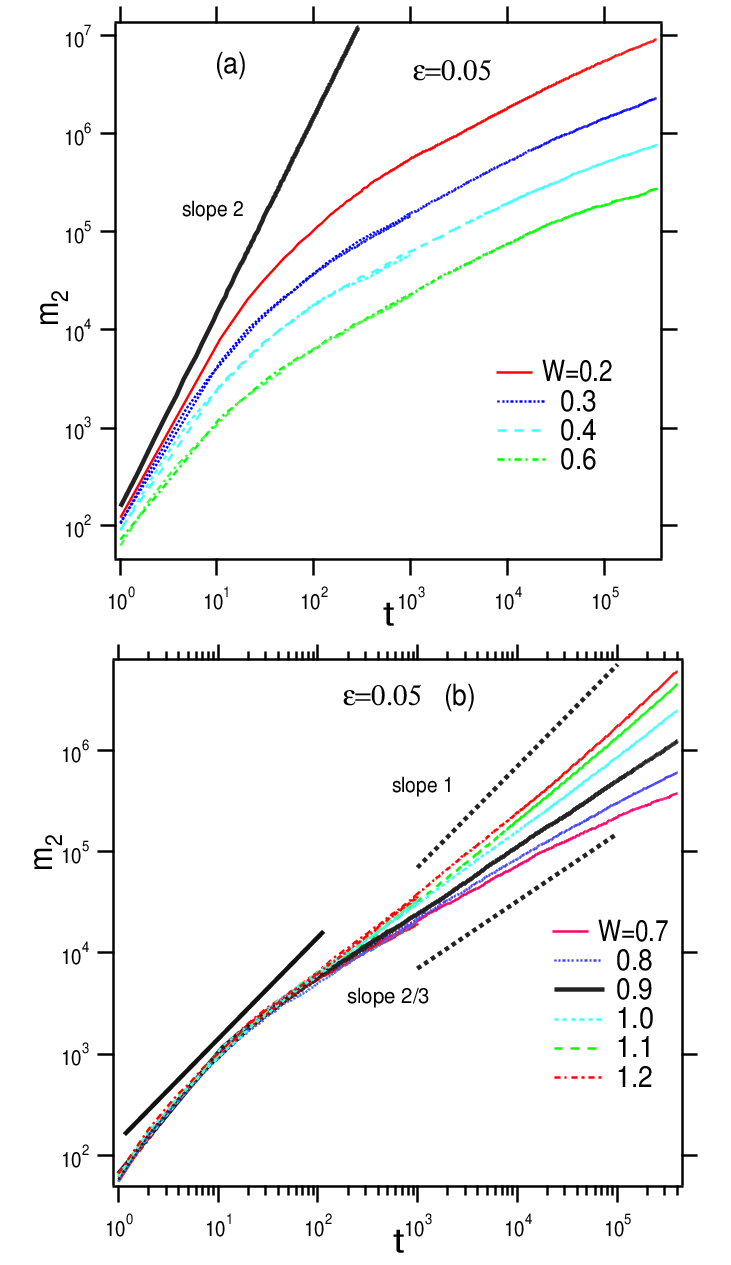}
\caption{(Color online)
The double-logarithmic plots of 
$m_2(t)$ as a function of $t$ for $M=2$ with  $\eps=0.05$ 
including the short time region.
(a)$W<W^*$, (b)$W>W^*$.
}
\label{fig:c2c3-e005-start}
\end{center}
\end{figure}

 Figures \ref{fig:c2c3-e005-start}(a) and (b) show the $m_2(t)$ 
divided into two regions,  $W<W^{*}$ and $W>W^{*}$, respectively, 
in the dichromatically perturbed AM with $\eps=0.05$.
This case corresponds to $L_3$  in Fig. 1.
In the case of $W<W^{*}$, the MSD increases as the $W$ decreases, 
but the shape of the curve is the same and no transition to the delocalization is seen.
On the other hand, in the case of $W>W^{*}$,  the LDT occurs 
around $W_c \simeq0.9$ and the MSD shows subdiffusive behavior $m_2 \sim t^{2/3}$ 
at the critical point.
It may seem strange that the localized quantum state delocalizes 
as the disorder width $W$ grows, 
but this is  one of the features of the perturbed AM with 
the characteristic value $W^{*}(<W_c)$.
Of course, it can be seen that when the $W$ is made larger ($W>>W_c$), 
the wavepacket spreads closer to the normal diffusion.
We can also see how the growth of MSD changes with the value of $W$.
In the region $W <W^*$, the ballistic increase is remarkable 
at the very short-time stage ($t<10$), 
but in the $W>W^*$, after the ballistic growth of the short time ($t<10$), 
it shows the diffusive spread after the time domain.

\begin{figure}[htbp]
\begin{center}
\includegraphics[width=6.0cm]{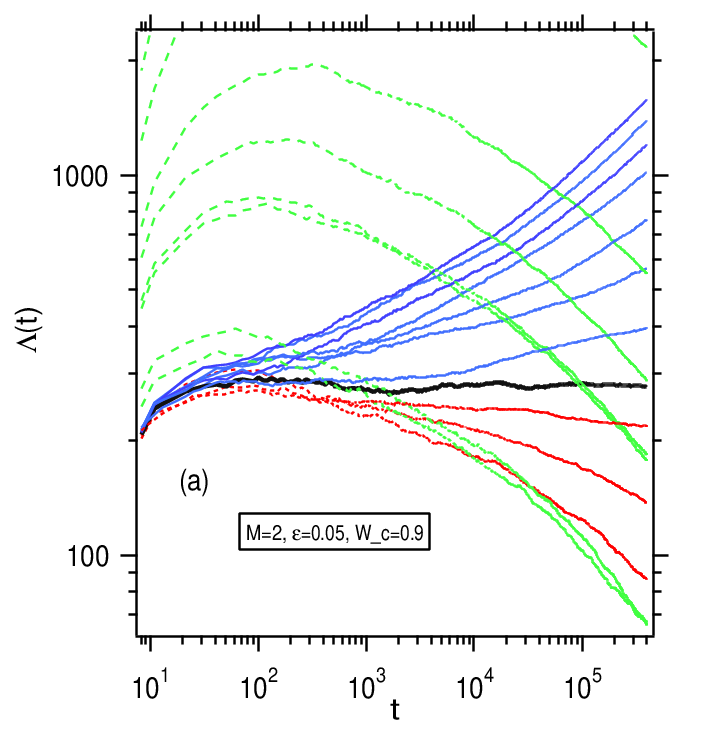}
\hspace{5mm}
\includegraphics[width=6.0cm]{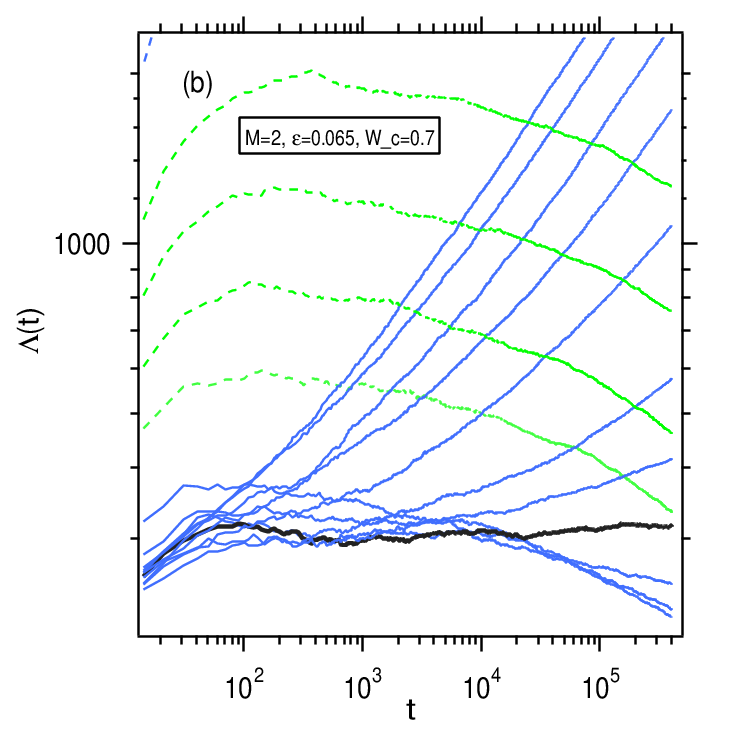}
\caption{(Color online)
The double-logarithmic plots of scaled $\Lambda(W,t)=m_2(t)/t^{2/3}$ 
 as a function of time in the dichromatically perturbed AM ($M=2$). 
(a)The case with $\eps=0.05$
for different values of the disorder strength $W$ along the $L_3$ in Fig.1, 
 where the critical case $W_c \simeq 0.9$ is shown by the thick black line.
(b)The case with $\eps=0.065$, where the critical case $W_c \simeq 0.7$
is shown by the thick black line.
The trumpet-shaped behavior of the $\Lambda(W,t)$ can be 
observed in the panel (a) because $W^*<W_c$.
}
\label{fig:c2-scaled-MSD}
\end{center}
\end{figure}


In Fig.\ref{fig:c2-scaled-MSD}(a), in order to see the transition phenomenon 
along with the L3 in the case of $\eps=0.05$ in which $W_c$ is larger than $W^*$ ($W_c>W^*$), 
the time-dependence of scaled MSD, $\Lambda(t,W) = \frac{m_2(t)}{t^{2/3}}$
is plotted for various $W$s.
When $W$ is increased the curve moves from top to bottom and rises again 
(denoted by green $\to$ red $\to$ blue) toward the LDT.
In the region $W<W^{*}$, even if the $W$ is increased, 
the MSD decreases, but there is no transition yet, 
and when $W$ further increases from the  $W \simeq W^{*}$
 at the bottom,  the LDT occurs at $W_c$.
We can see the trumpet-shaped change of the scaling function $\Lambda(t)$
around the critical region $W\simeq W_c$.
It should be noted that the two types of the decreasing curve 
(denoted by green and red lines, respectively)
are different in the shape (quality).
Accordingly, it follows that they cannot be overlapped in the one-parameter scaling analysis.

On the other hand, Fig.\ref{fig:c2-scaled-MSD}(b) shows $m_2(t)$
 along the line $L_2$ in the perturbed AM with $\eps=0.065$ with increasing $W$ 
in the case of $W_c < W^{*}$.
The LDT occurs  after the MSD decreases with the increase of $W$, 
but there is almost no trumpet-shaped lower critical region.

\subsection{Scaling analysis of  the LDT}
\label{app:scaling-AM}
Similar to the case of the LDT with increasing $\eps$ in the previous paper, 
here the critical exponent of the localization length 
can be determined by performing finite-time scaling 
analysis in the critical region of the LDT with changing $W$.
This corresponds to the finite-size scaling analysis 
for Anderson transition in the higher-dimensional random system 
\cite{markos06,garcia07,slevin14,tarquini17}.

\begin{figure}[htbp]
\begin{center}
\includegraphics[width=7.5cm]{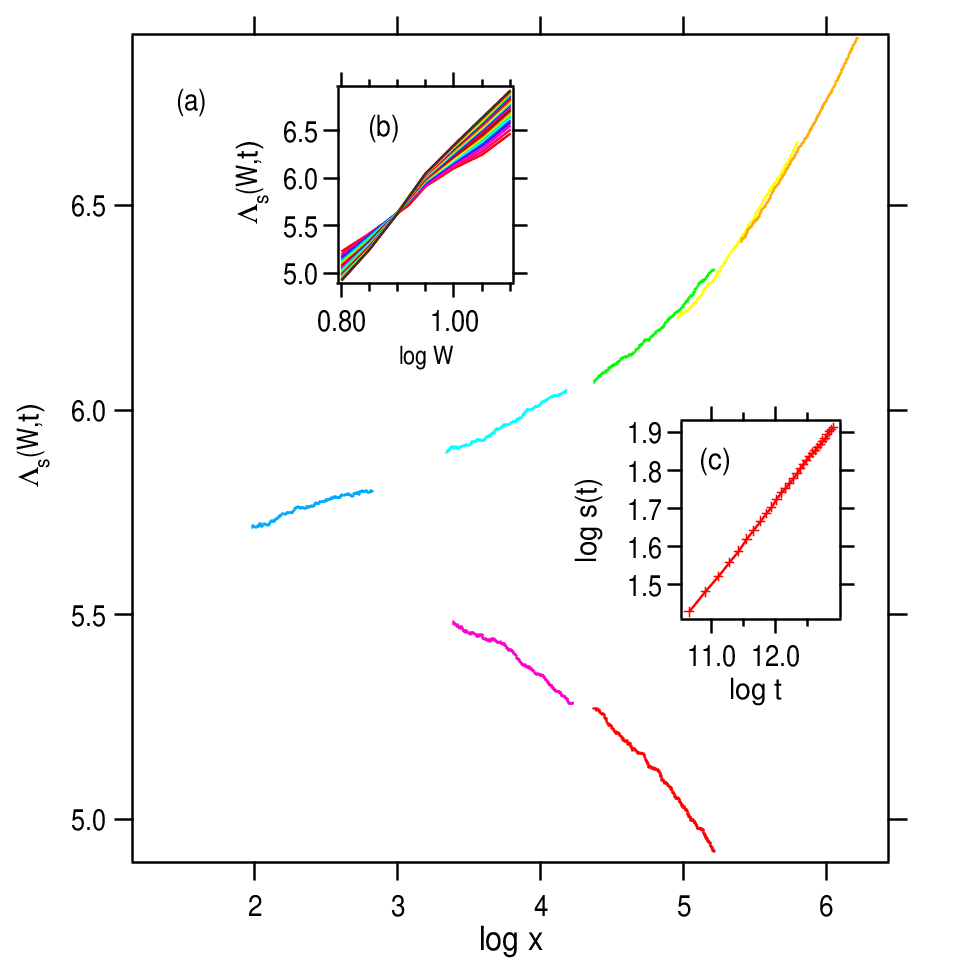}
\caption{(Color online)
The results of the critical scaling analysis for 
dichromatically perturbed AM ($M=2$) with $\eps=0.05$.
(a)The scaled MSD $\Lambda_s(W,t)$ with $\alpha=0.65$ 
as a function of $x=\xi_0|W_c-W|^{-\nu}t^{\alpha/2\nu}$
for some values of $W$. 
(b)The same scaled MSD $\Lambda_s(W,t)$ as a function of  $W$
for some pick-up time. 
The crossing point is $W_c \simeq 0.9$.
(c)$s(t)=\log(\Lambda(W,t)/\Lambda(W_c,t))/(W_c-W)$ as a 
function of $t$.
The critical exponent $\nu \simeq 1.48$
is determined by a scaling relation Eq.(\ref{eq:critical-exponent}) 
by the least-square fit for data in  (c).
}
\label{fig:c2-e005-nu}
\end{center}
\end{figure}

Figure \ref{fig:c2-e005-nu} displays the results of the  finite-time scaling 
analysis for dichromatically perturbed AM ($M=2$) with $\eps=0.05$ corresponding to Fig.6.
We choose the following quantity as a scaling variable
\beq
\Lambda_s(W,t)=\log \left[ \frac{m_2(t)}{t^\alpha} \right].
\eeq
For $W > W_c$, 
the  $\Lambda_s(t)$ increases and the wave packet delocalizes with time.
On the contrary, for  $W < W_c$,  $\Lambda_s(t)$ decreases with time and the
wave packet turns to the localization.
In the vicinity of the LDT, for $\Lambda_s(t)$, OPST
 is assumed with  localization length $\xi_s(W)$ as the parameter.
Then, $\Lambda_s(W,t)$ can be expressed as,
\beq
\Lambda_s(W,t) &=& F(x),
\label{eq:real-scale-0}
\eeq
where 
\beq
x=|W_c-W|t^{\alpha/2\nu}.
\eeq
$F(x)$ is a differentiable scaling function and the $\alpha$ is the diffusion index.
Figure \ref{fig:c2-e005-nu}(b) shows the plot of $\Lambda_s(t)$ 
as a function of $W$ at several times $t$, and 
it can be seen that this intersects at the critical point $W_c$.
In addition, Fig.\ref{fig:c2-e005-nu}(c) shows the plot of 
\beq
s(t)&=&\frac{\Lambda_s(W,t)-\Lambda_s(W_c,t)}{|W_c-W|} \\
  &\propto&  t^{\alpha/2\nu}.
\label{eq:critical-exponent}
\eeq
as a function of $t$, and the critical exponent $\nu=1.48$ 
of the LDT is determined by best fitting this slope which 
is a similar result obtained in the previous paper 
for the LDT \cite{yamada19a}. 

In Fig.\ref{fig:c2-e005-nu}(a), we plot $\Lambda_s$ 
as a function of $x=t^{\alpha/2}/\xi_s(W)$ 
for different values of $W$ by using the obtained the critical exponent $\nu$.
It is well scaled and demonstrates the validity of the OPS. 
The upper and lower curves represent the delocalized and localized branches
of the scaling function, respectively.
The establishment of OPS shows 
the equivalence of the time change  and the parameter change 
in the  $m_2(t,\eps,W)$.

Around the critical point of the LDT, the localization length $\xi_s$
 is supposed to diverge as
\beq
\xi_s \simeq \xi_0 |W_c-W|^{-\nu}
\eeq
at $W=W_c$.  
($\xi_s$ depends on the number of modes $M$, but the subscript $M$ is omitted.)

\section{Delocalized states and Normal Diffusion}
\label{sec:delocalized-state}
In this section, we evaluate how the delocalized states spread due to 
the changes in $W$ and $\eps(>\eps_c)$.
Is this true that if $\eps>\eps_c$ the delocalized states 
asymptotically approach the normal diffusion as $t \to \infty$?
Little is known about the models of quantum systems 
in which normal diffusion occurs without any stochastic fluctuation \cite{yamada12}.

\subsection{Delocalized states}
\begin{figure}[htbp]
\begin{center}
\includegraphics[width=6.1cm]{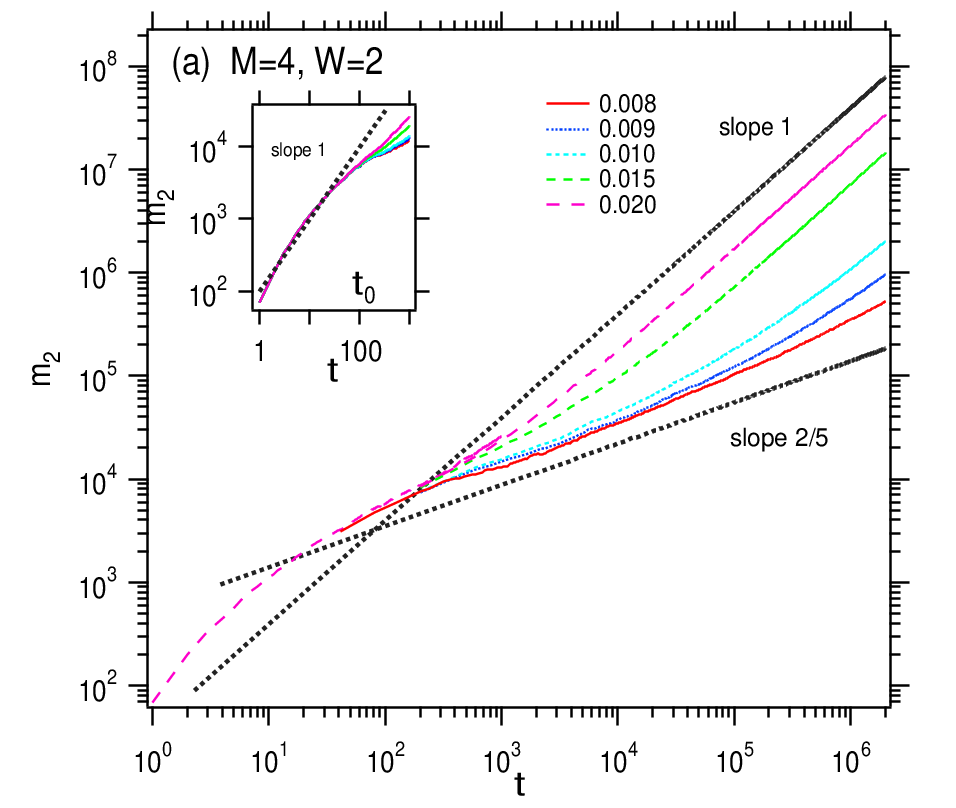}
\hspace{3mm}
\includegraphics[width=6.1cm]{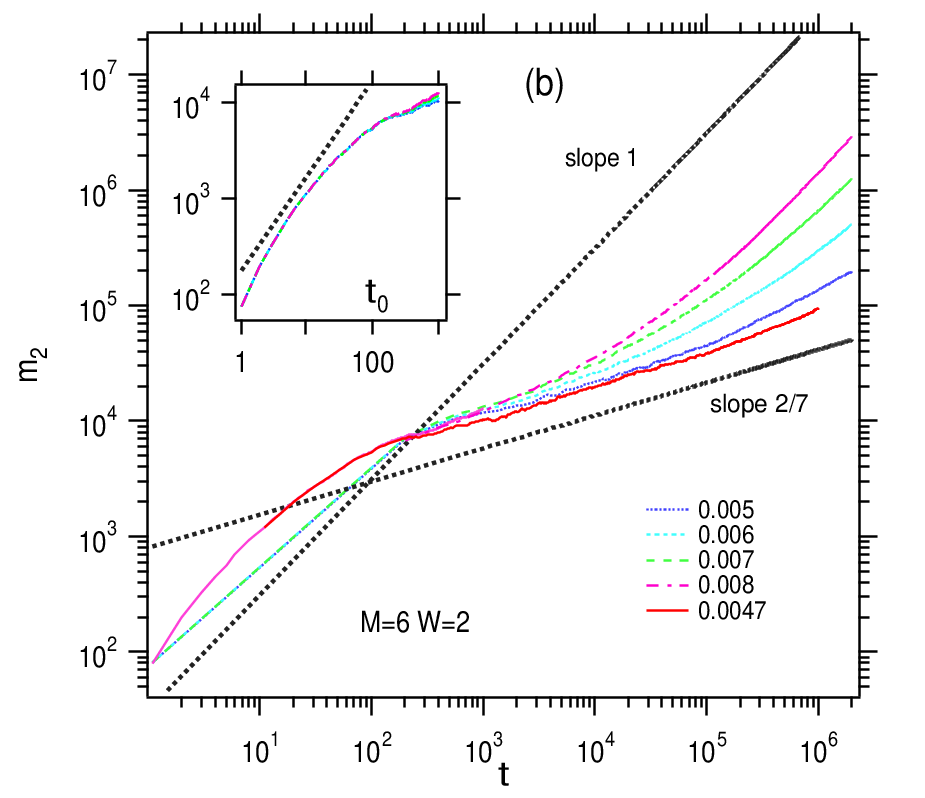}
\caption{
(Color online)
The double-logarithmic plots of 
$m_2(t)$ as a function of $t$ for different values of $\eps$ in 
the polychromatically perturbed AM with  $W=2$ including the detailed short time region.
 (a)$M=4$ and (b)$M=6$.
The insets are the enlarged view of the short-time region $t<10^3$.
}
\label{fig:W2-c4c6-long}
\end{center}
\end{figure}

Figure \ref{fig:W2-c4c6-long} shows the long-time behavior of the MSD 
for  $\eps>\eps_c$ in the polychromatically perturbed AM ($M=4,6$).
It seems that regardless of $M$ the MSD approaches the 
normal diffusion as $m_2 \sim t^1$ when $\eps$ is large.
It can be expected that the normal diffusion 
occurs in a longer time even if $\eps$ is relatively small if $\eps>\eps_c$
because the scaling curve for various $\eps$s neatly fit on one curve.
The inset is an enlarged view in a short-time domain ($t <10^2(\equiv t_0$) )
before the perturbation starts to work.
Here, it can be seen that  since $W=2(>W^*)$ the perturbation works 
after passing through the diffusive time domain ($t<t_0$).

\subsection{$W-$dependence of the diffusion coefficient}
\begin{figure}[htbp]
\begin{center}
\includegraphics[width=7.0cm]{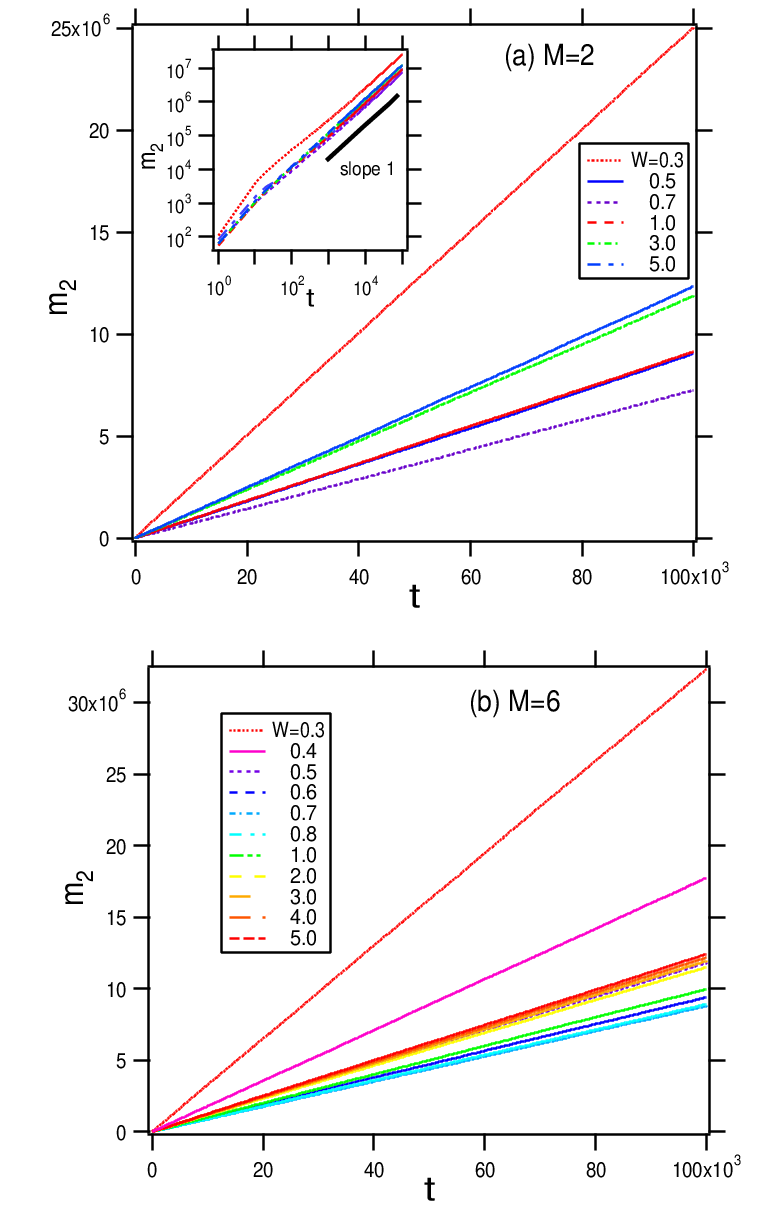}
\caption{(Color online)
The plots of 
$m_2(t)$ as a function of $t$ for different values of $W$ in 
the polychromatically perturbed AM  with  $\eps=0.2$. 
 (a)$M=2$ and (b)$M=6$.
Note that the axes are in the real scale. 
The inset of the panel (a) is in the logarithmic scale. 
The black dotted line shows $m_2(t) \propto t^{1}$
for reference.
}
\label{fig:A-model-msd}
\end{center}
\end{figure}

\begin{figure}[htbp]
\begin{center}
\includegraphics[width=7.0cm]{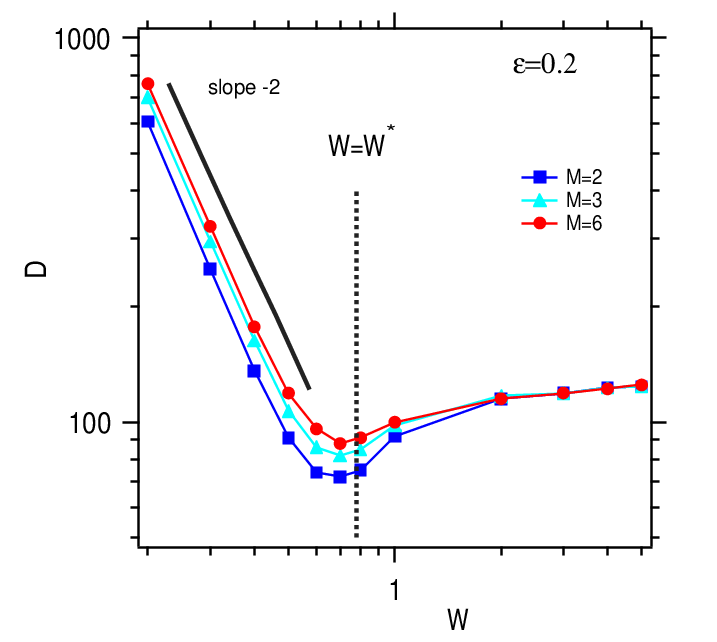}
\caption{(Color online)
The diffusion coefficient $D$ of the quantum diffusion 
 as a function of $W$ 
in the polychromatically perturbed AM with $\eps=0.2$ of  $M=2,3,6$.
Note that the axes are in the logarithmic scale. 
$D \propto W^{-2}$ and $W^*=0.78$ are shown by the black line and 
the black dotted lines, respectively, for reference.
}
\label{fig:A-model-diff}
\end{center}
\end{figure}

As shown in Fig.\ref{fig:c2c3-e005-start}, it can be expected that 
$m_2(t)$ asymptotically approaches to the normal diffusion
as the  $W(>W_c)$ increases for the fixed $\eps$. 
 Indeed, Fig.\ref{fig:A-model-msd} shows the time-dependence of the MSD
by changing $W$ in the case of the polychromatically perturbed AM
with $\eps=0.2(>>\eps_c)$.
As shown in Fig.\ref{fig:A-model-msd}(a), 
the $m_2(t)$ behaves the normal diffusion, 
\beq
m_2(t) \simeq Dt,
\eeq
for $M=2$, where $D$ is the diffusion coefficient.
As shown in Fig.\ref{fig:A-model-msd}(b), 
similar results are confirmed even for the case of $M=6$.
Figure  \ref{fig:A-model-diff} shows the $W-$dependence of 
the diffusion coefficient numerically estimated.
Regardless of the number of colors $M$, in the region $W<<W^*$ it behaves
\beq
 D \propto \frac{1}{W^2}.
\eeq
However, when  the $W$ increases, it does not decrease monotonously, 
but it shows the minimum of $D$ around $W^*$ and gradually increases towards
the constant value and it becomes 
\beq
 D \sim const.
\eeq
for $W>>W^*$.
This behavior can be inferred from the Maryland transform as follows.
For $W \simeq W^*$, the effect of the randomness of the diagonal term saturates, 
and the range of the random hopping term still increases even for $W>W^*$
when the $\eps=0.2$.
This tendency does not depend on the number of colors $M$ after the LDT.

%

The above behavior of the diffusion coefficient 
for the weak disorder region ($W<W^*$)  
can be explained from the $W-$dependence of the localization length 
in the localized side.
First, the $W-$dependence of the localization length is assumed 
to be $\xi(W) \propto 1/W^2$ for $W<<1$.
As seen in Table 1, for $\eps>\eps_c$, 
the MSD changes from the ballistic motion to the diffusion as 
the time elapses.
If the duration (characteristic time) of this ballistic motion is $\tau$, 
it can be estimated as 
\beq
\tau \simeq \frac{\xi(W)}{2\pi}.
\eeq
This is equivalent to considering the time evolution of the wave packet 
in Brownian motion, assuming that memory is lost at $\tau$, 
for the weak disorder region ($W<W^*$).
By using the characteristic time $\tau$, the time-dependence of 
$m_2(t)$ is expressed as
\beq
 m_2(t) \sim \xi^2 \frac{t}{\tau},
\eeq
if we consider $\xi$ spreads every time $\tau=\xi/2\pi$.
Then if the $\eps$ is fixed, the diffusion coefficient becomes
\beq
D &=& \lim_{t \to \infty}\frac{m_2(t)}{t} = \frac{\xi^2}{\tau} \\
& \sim &  \frac{1}{W^2}.
\eeq
As a result, the $W-$dependence of the localization length
 propagates to that of the diffusion coefficient.

\subsection{$\eps-$dependence of the diffusion coefficient}

\begin{figure}[htbp]
\begin{center}
\includegraphics[width=7.0cm]{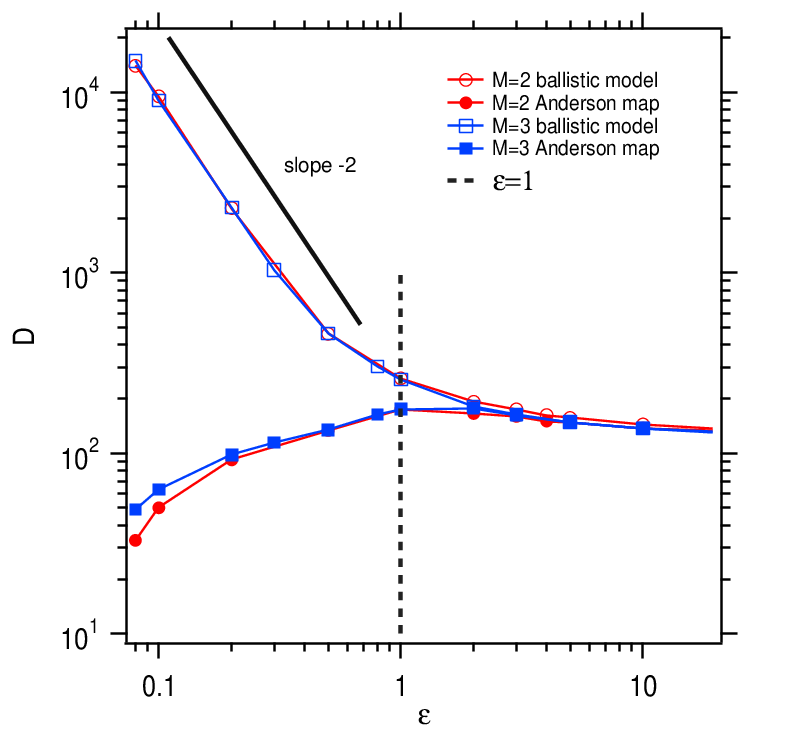}
\caption{(Color online)
The diffusion coefficient $D$ of the quantum diffusion 
 as a function of  $\eps$ 
in the polychromatically perturbed AM with $W=1$ of  $M=2,3$.
The corresponding results for the ballistic model are also provided.
Note that the axes are in the logarithmic scale. 
$D \propto \eps^{-2}$ and $\eps=1$ are shown by the black line and 
the black dotted lines, respectively, for reference.
}
\label{fig:A-eps-difcof-1}
\end{center}
\end{figure}

Next, let's examine the $\eps-$dependence of the diffusion coefficient 
of the delocalized states.
The $\eps-$dependence at $\eps>\eps_c$ for $M=2,3$, fixed at $W=1$, 
is shown in Fig.\ref{fig:A-eps-difcof-1}.
It can be seen that the diffusion coefficient of the delocalized state 
increases gradually with increasing $\eps$, and saturates beyond $\eps \simeq 1$.
This tendency also does not depend on the number of colors 
if the LDT occurs when $M \geq 2$.
What does this value $\eps \simeq 1$ mean?
Its meaning can be understood by replacing 
 the time-dependent part $f(t)$ of the potential with
\beq
g(t)=\left[\frac{\eps}{\sqrt{M}} \sum_i^M\cos(\omega_i t + \theta_i)\right].
\label{eq:gt}
\eeq
This case can be called a ballistic model because it show ballistic spreading 
$m_2 \sim t^2$ 
in the unperturbed case ($\eps=0$).
Actually, the $\eps-$dependence of the diffusion coefficient in the ballistic model is also plotted 
in Fig.\ref{fig:A-eps-difcof-1}.
It shows $D \propto \eps^{-2}$ for $\eps <1$ and 
it gradually decreases towards  a certain value 
when the perturbation strength $\eps$ increases.
This is a reasonable result 
because the effect of the presence of $"1"$ that gives localization at the case of $f(t)$
relatively  weakens, and the time-varying term becomes more dominant. 
As a result, for $\eps>>1$ it can be seen that the value of the diffusion coefficient 
in AM is asymptotic to that in the ballistic model.

See appendix \ref{app:B-model} for the details of the 
$W-$dependence of the normal diffusion in the ballistic model.

%
%

\section{Summary and discussion}
We investigated the localization-delocalization transition (LDT) of 
the AM (Anderson map) which 
are dynamically perturbed by polychromatically
periodic oscillations for the initially localized quantum wave packet.

The $W-$dependence and $\eps-$dependence of the localization length (LL)
in the completely localized region have characteristics similar to 
those of the monochromatically perturbed AM.
Since the characteristic value $W^*$ exists, the MSD shows a peculiar change 
for the change of $W$, but the critical behavior of the LDT 
for changing $W$ is similar to that in the LDT for changing $\eps$ around the
critical point.
In the region $W^*<W_c$, 
the shape of the scaling function changes, 
but the localization remains the same at $W=W^*$
when $W$ increases, and 
 the increasing $W$ causes the LDT at a certain value $W=W_c$.

We also studied the delocalized states for $\eps>\eps_c$.
The $W-$dependence of the diffusion coefficient of the delocalized states 
decreases in the form $D \propto W^{-2} $ for the region $W<W^*$.
The $D$ gradually increases towards $D \simeq const$ for $W>>W^*$
after it becomes the minimum diffusion coefficient around $W\simeq W^*$.

Roughly speaking, since the time-continuous system 
corresponds to $W^* \to \infty$ in the AM, 
it is expected that there is no LDT along the line $L_2$ and $L_3$ in Fig.1, 
 only LDT corresponding to 
 the region  $W <W^*$ along $L_1$ will be observed in the time-continuous Anderson 
model  with the time-quasiperiodic perturbation \cite{yamada99,yamada02,yamada20}.




\appendix

\section{Critical phenomena of LDT and the localization length} 
\label{app:LL-connect}
In this appendix, we consider the relation of the localization length $\xi(\eps)$
obtained by direct calculation for $\eps <<\eps_c$
and the LL by finite-time scaling analysis around $\eps \simeq \eps_c$
in the  polychromatically perturbed AM  $W=0.5$. 
We consider the LDT along the line $L_1$
but the same argument holds qualitatively for cases with the other parameters.
The localization length $\xi_{s}^{(M)}(\eps)$ obtained indirectly 
from the critical scaling analysis 
around the critical value $\eps_c^{(M)}$  is expressed,
\beq
\xi_s^{(M)}(\eps) = \xi_0^{(M)} |\eps_c^{(M)}-\eps|^{-\nu^{(M)}}, 
\label{eq:xi-eps-w3}
\eeq
where $\xi_{s}^{(M)}$, $\eps_{c}^{(M)}$  and $\nu^{(M)}$ are
the localization length , critical strength and the critical exponent of the LDT, respectively.
$A_0^{(M)}(W)=\xi_0 (\eps_c^{(M)})^{\nu}$ can be determined by the LL of
the unperturbed case $\xi_0=\xi(\eps=0)$.
The order of the localization length for the different  $M$ satisfies 
\beq
 \xi_{s}^{(2)}(\eps)< \xi_{s}^{(3)}(\eps) < \xi_{s}^{(4)}(\eps) ...
\eeq
In addition, the following relations of the critical strength and critical exponent holeds,
\beq
\eps_{c}^{(2)} & >& \eps_{c}^{(3)} >  \eps_{c}^{(4)}  > ...., \\
\nu^{(2)} & >& \nu^{(3)}  > \nu^{(4)} > .....
\eeq
It is established from the results and theory of the finite-time scaling analysis.
In other words, considering the $M-$dependence of this transition, 
the critical value $\eps_c$ becomes smaller and the critical exponent $\nu$ 
also becomes smaller as the number of modes $M$ increases, 
and the divergence of the LL around the critical value becomes mild as the $M$ increases.

\begin{figure}[htbp]
\begin{center}
\includegraphics[width=8cm]{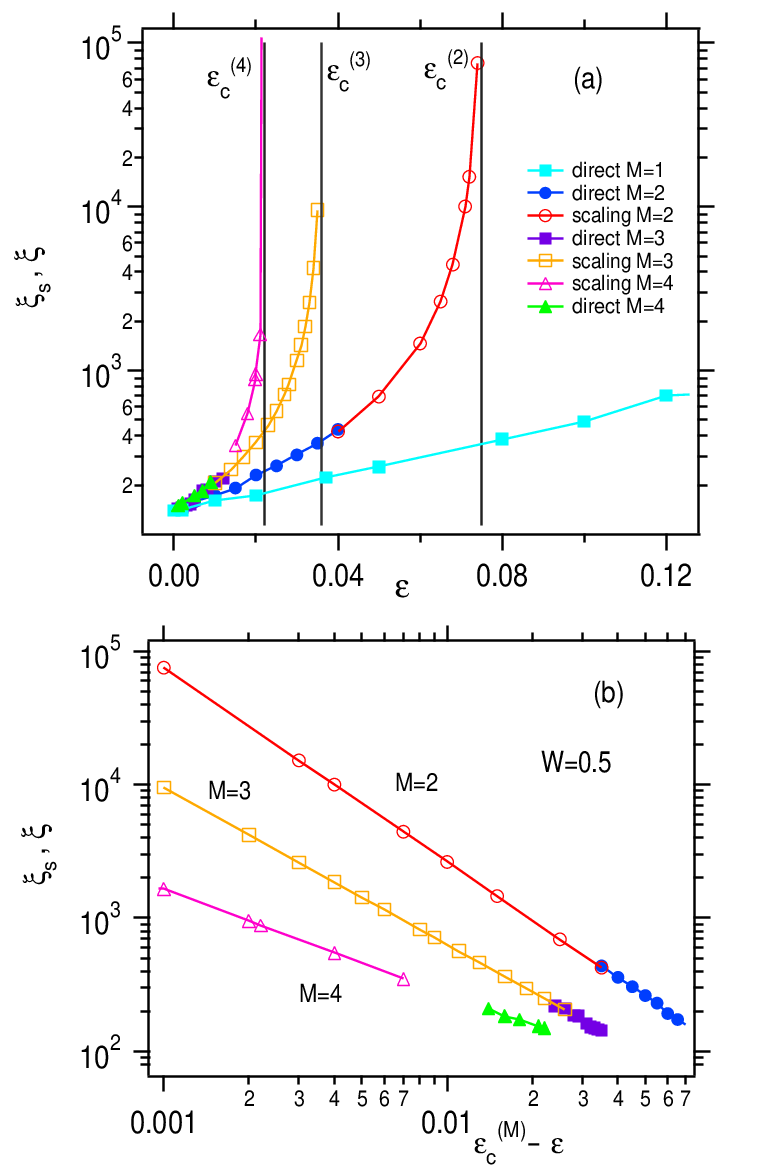}
\caption{(Color online)
(a)The localization length $\xi(\eps)$, $\xi_s(\eps)$as a function of $\eps$
for perturbed AM ($M=1,2,3,4$) with $W=0.5$.
The filled symbols denote the numerical data obtained by 
$\xi(\eps)=\sqrt{m_2(t\to\infty)}$ in the long-time limit.
The open symbols indicate the localization length $\xi_s(\eps)$ 
obtained by the OPST in the critical region. 
Note that the vertical axis is in the logarithmic scale.
(b) $\xi_n(\eps)$ and $\xi_s(\eps)$ as a function of $\eps_c^{(M)}-\eps$.
}
\label{fig:c2c3c4-LL-eps}
\end{center}
\end{figure}

In Fig.\ref{fig:c2c3c4-LL-eps}, we compare  the localization length  
$\xi_s^{(M)}(\eps)$ decided indirectly by OPST 
in the critical region $\eps\simeq\eps_c$ with $\xi^{(M)}(\eps)$ decided directly
by the saturated MSD data which are precisely calculated for $\eps$'s much less 
than the critical region.
The $\eps-$dependence of these two localization lengths, 
$\xi_s^{(M)}(\eps)$, $\xi^{(M)}(\eps)$,
 seem to connect continuously, which 
implies unexpected wideness of the critical region in which the OPST works
\cite{footnote:d-dds}. 

On the other hand, as seen in the main text, even if $M=2$ and $M=3$, 
the LL increases exponentially as $\eps$ increases at least for $\eps<<\eps_c$,
according to the Eq.(\ref{eq:AM-last}). 
Furthermore, it seems that 
when $\eps$ increases, the LL increases with the increasing of the $\eps$
 more strongly than the exponential growth.
This is natural because $\eps$ grows closer to $\eps_c$ and 
leads to the divergence of the LL at $\eps=\eps_c$.


\section{Diffusive characteristics of Ballistic model}
\label{app:B-model}

Here we consider a model without the localization 
by replacing the time-dependent part $f(t)$ 
of Eq.(\ref{eq:perterbation}) with $g(t)$ of Eq.(\ref{eq:gt}).
The system describes the free particle scattered by the quasiperiodically
oscillating irregular potential.
This model is also interesting as it is connected with the problem of 
ballistic electrons scattered by dynamical impurities.
We refer to this model as the ballistic model  in this paper.
The AM with  the time-dependent part $f(t)$ approaches the ballistic model
in a limit $\eps \to \infty$, and the system has complete parameter dependence
 in the form of $\eps W$.

In the ballistic model, a simple interpretation by Maryland transform is possible.
The diagonal term is a constant value that does not depend on the site.
If $W=0$ (equivalent to $M=0$), the hopping terms are also constant, 
and the Hamiltonian describes the tight-binding system with  periodic potential.
Accordingly, the dynamics of the unperturbed case exhibits the ballistic spread 
instead of the localization such as, 
\beq
m_2(t) \sim t^2.
\eeq
Such a ballistic motion is suppressed and changes into aother kind of 
motion by introducing the dynamically oscillating part
($W\neq 0$, $\eps \neq 0$).

\begin{figure}[htbp]
\begin{center}
\includegraphics[width=8.8cm]{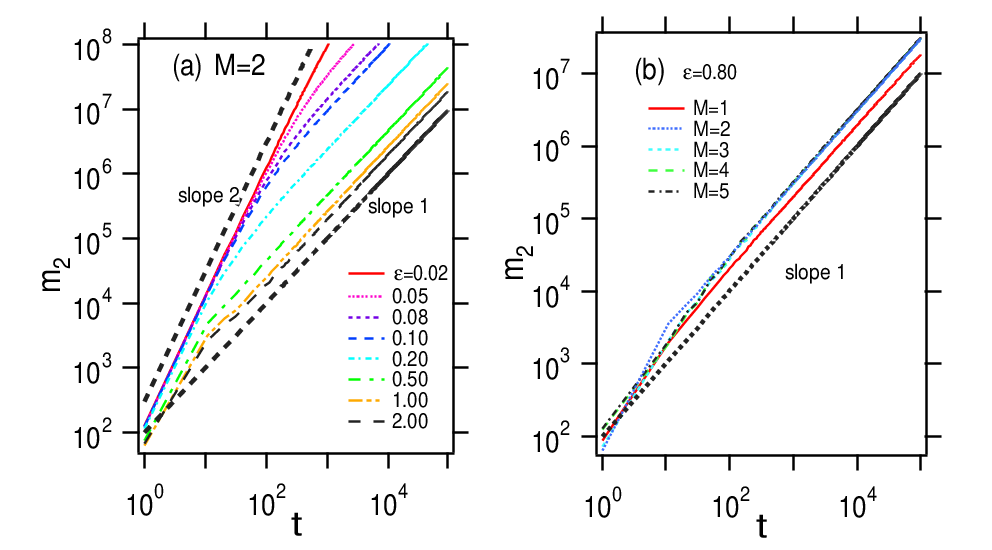}
\caption{(Color online)
The double-logarithmic plots of 
$m_2(t)$ as a function of $t$ for different values of $\eps$ in 
the polychromatically perturbed ballistic model $W=1$. 
 (a)$M=2$ with various values of $\eps$s, and (b)$M=1\sim5$ with  $\eps=0.8$.
The black dotted lines show $m_2(t) \propto t^{2}$ and $m_2(t) \propto t^{1}$
for reference.
}
\label{fig:ballistic-msd}
\end{center}
\end{figure}

Figure \ref{fig:ballistic-msd} shows the time-dependence of the MSD 
with changing $\eps$ and $M$ as the perturbation is applied.
As shown in Fig.\ref{fig:ballistic-msd} (a), 
the MSD changes from $m_2(t) \sim t^2$ to $m_2(t) \sim t^1$ when the perturbation tern on.
Similarly, as shown in Fig.\ref{fig:ballistic-msd}(b), 
there is no subdiffusion
even in the cases of $M=1\sim5$, and it behaves normal diffusion, 
$m_2(t) \simeq Dt$ for $t>>1$, where $D$ is the diffusion coefficient.

\begin{figure}[htbp]
\begin{center}
\includegraphics[width=7.0cm]{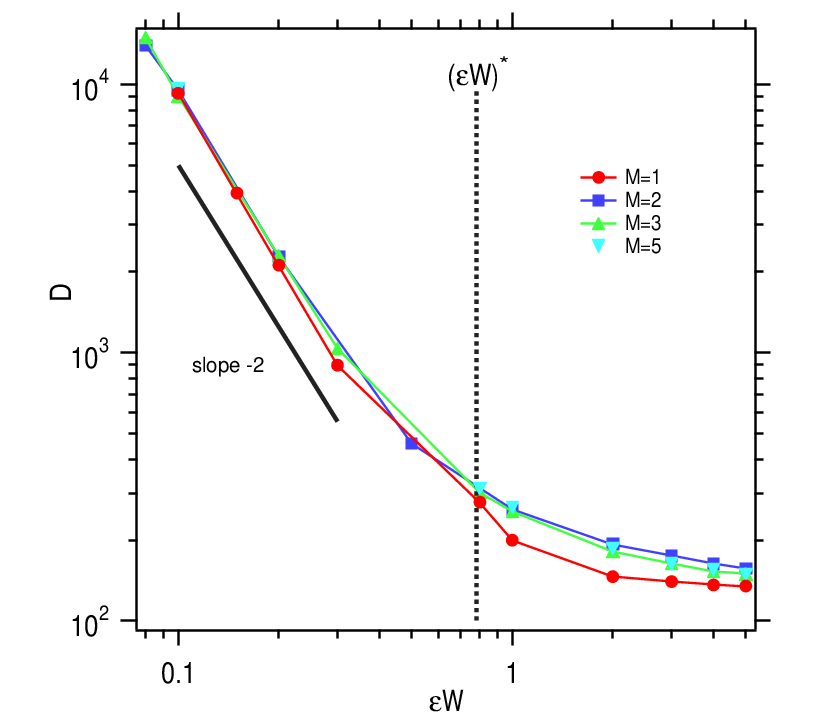}
\caption{(Color online)
The diffusion coefficient $D$ of the quantum diffusion 
 as a function of $\eps W$ 
in the polychromatically perturbed ballistic model
($M=1\sim5$).
Note that the axes are in the logarithmic scale. 
$(\eps W)^*=0.78$ is shown by the black dotted lines for reference.
}
\label{fig:B-model-diff}
\end{center}
\end{figure}

The $(\eps W)-$dependence of the diffusion coefficient $D$ is shown in 
Fig.\ref{fig:B-model-diff}. 
It can be seen that when the parameter  $\eps W$ becomes larger than $(\eps W)^*=0.78$
, the $D$ gradually approaches a certain value.
It follows that  even in the monochromatically perturbed AM, there is no localization 
and normal diffusion is achieved.
These phenomena can also be interpreted based on the Maryland transform.
For $\eps W>0$, it becomes diffusive  by the hopping term including the disorder.
When the $W$ farther increases,  the diffusion is suppressed due to the disorder
and the diffusion coefficient is reduced.
Even if $W$ is further increased, the disorder effect seems to be saturated
because  the off-diagonal term is also tangent-type.

\section*{Acknowledgments}
This work is partly supported by Japanese people's tax via JPSJ KAKENHI 15H03701,
and the authors would like to acknowledge them.
They are also very grateful to Dr. T.Tsuji and  Koike memorial
house for using the facilities during this study.



\begin{thebibliography}{00}
\bibitem{footnote:d-dds}
This is also true for the Anderson transition
 in the three-dimensional disordered system.
Mathematically, there is a proof of the existence of low-energy eigenstates
 characterized by the localization length 
even in the higher-dimensional random systems.
Something is interesting about how 
the localization length connect to that numerically obtained 
by the finite-size scaling?
\end{thebibliography}
\end{document}